\documentclass[sn-mathphys]{sn-jnl}
\jyear{2025}
\theoremstyle{thmstyleone}

\theoremstyle{thmstyletwo}

\theoremstyle{thmstylethree}

\usepackage{siunitx}
\usepackage{eso-pic}
\AddToShipoutPictureFG{
  \AtPageLowerLeft{
    \put(\LenToUnit{1.8cm},\LenToUnit{1.5cm}){\footnotesize Accepted by RDTM. https://doi.org/10.1007/s41605-025-00599-5}
  }
}

\begin{document}

\title[ ]{Characterization of SiPMs at 40~K for neutrino coherent detection based on pure CsI}

\author[1,2,3]{\fnm{Tao} \sur{Liu}}
\author*[3,4]{\fnm{Xilei} \sur{Sun}}\email{sunxl@ihep.ac.cn}
\author*[1,2]{\fnm{Fengjiao} \sur{Luo}}\email{luofengjiao@usc.edu.cn}
\author[3,4]{\fnm{Jingbo} \sur{Ye}}
\author[1,2]{\fnm{Bo} \sur{Zheng}}
\author[3]{\fnm{Cong} \sur{Guo}}
\author[3]{\fnm{Zhilong} \sur{Hou}}
\author[3]{\fnm{Rongbin} \sur{Zhou}}
\author[1]{\fnm{Aiqin} \sur{Gao}}
\author[3]{\fnm{Lei} \sur{Cao}}
\author[1,2]{\fnm{Bo} \sur{Zhang}}
\author[1,2]{\fnm{Sijia} \sur{Han}}

\affil[1]{\orgdiv{School of nuclear science and technology}, \orgname{University of South China}, \orgaddress{\city{Hengyang}, \postcode{421001}, \state{Hunan}, \country{China}}}

\affil[2]{\orgdiv{Key Laboratory of Advanced Nuclear Energy Design and Safety}, \orgname{Ministry of Education}, \orgaddress{\city{Hengyang}, \postcode{421001}, \state{Hunan}, \country{China}}}

\affil[3]{\orgdiv{Institute of High Energy Physics}, \orgname{Chinese Academy of Sciences}, \orgaddress{\city{Beijing}, \postcode{100049}, \country{China}}}

\affil[4]{\orgname{State Key Laboratory of Particle Detection and Electronics}, \orgaddress{\city{Beijing}, \postcode{100049}, \country{China}}}

\abstract{Silicon photomultiplier (SiPM), as the core photoelectric sensor for coherent neutrino detection in low-temperature pure CsI, its working performance directly determines the measurement accuracy of the scintillator light yield. Our previous research has fully demonstrated the performance of pure CsI at liquid nitrogen temperature. More intriguingly, its performance is expected to be even better at 40~K. However, the performance characteristics of SiPM in the 40~K temperature range still remain to be explored. In this study, a self-developed adjustable temperature control system ranging from 30~K to 293~K was built to investigate the key performance parameters of SiPM at different temperatures, such as single photoelectron spectrum, gain, breakdown voltage, dark count rate, after-pulse, internal crosstalk, and single photoelectron resolution. Special emphasis was placed on examining the key performance parameters of SiPM in the 40~K temperature range to evaluate its feasibility for light yield measurement in this temperature range. The results show that this study obtained the parameter variation trends and optimal working conditions of 3 types of SiPM at different temperatures, thereby improving the sensitivity of the detector. This research provides important technical support for low-temperature detection in neutrino physics experiments.}


\maketitle
\section{Introduction}\label{sec1}
Coherent Elastic Neutrino-Nucleus Scattering (CE$\nu$NS) is a new type of low-energy neutrino detection method. It was first experimentally confirmed by the COHERENT collaboration in 2017 \cite{1}, and this discovery came more than 40 years after the theoretical prediction \cite{2}. The coherent scattering cross-section is the largest among the known neutrino detection cross-sections, which is proportional to the square of the number of neutrons in the atomic nucleus \cite{3}. Its cross-section is more than two orders of magnitude higher than that of inverse beta decay (IBD) \cite{1}. The high cross-section means that the miniaturization of detectors can be realized, so coherent scattering experiments have become an emerging direction in neutrino experiments. As a weak neutral current, CE$\nu$NS has become a new tool for studying neutrino properties and nuclear physics. It is of great significance for the research on neutrino magnetic moments \cite{4,5}, non-standard neutrino interactions (NSI) \cite{6,7,8,9}, as well as probes of nuclear structure \cite{10,11,12}.

The COHERENT experiment uses CsI(Na) crystals with a Na concentration of 0.114 mol\% \cite{1,13}, which is about an order of magnitude higher than that of conventional CsI(Na) crystals. This can shorten the distance between the luminescent centers of Na ions in the crystal and increase the light yield of nuclear recoil luminescence. Even so, the photoelectron output level and quenching factor are still limited \cite{1,14,15}, and most excitons do not emit light due to thermal quenching. We propose a new detector scheme based on pure CsI at 77~K and SiPM readout, and the preliminary experiment has verified its feasibility \cite{16}. Low temperature effectively reduces the probability of exciton thermal quenching in CsI crystals and increases the probability of electrons being captured by holes to form excitons and emit light \cite{17,18}. At 40~K, the light yield is further increased, and the $\alpha$ excitation light yield is higher than that of $\gamma$ excitation \cite{19,20,21,22}.

The study uses SiPM with high photon detection efficiency as the photoelectric readout device, and the SiPM is coupled with the crystal through silicone oil. SiPMs need to operate at low temperatures of 40~K, but most of the test data provided by manufacturers is concentrated in the room temperature and above temperature range \cite{23}, and the performance data at low temperatures is still incomplete, which makes it difficult to design and optimize low-temperature detectors. Therefore, systematic research on the low-temperature performance of SiPMs is of great practical significance.

This study conducted performance tests on SiPMs from three manufacturers, namely Hamamatsu, New Device Laboratory (NDL), and Broadcom, over the temperature range from 30~K to 293~K, and systematically evaluated their application potential in low-temperature detectors. The experiment focused on the key performance indicators of SiPMs in a 40~K environment, including the variation trends of parameters such as gain, breakdown voltage (\(V_{bd}\)), dark count rate (DCR), after-pulse, internal crosstalk (iCT), and single photoelectron (SPE) resolution. This study recorded and analyzed the variation laws of the key parameters of these three types of SiPMs within the range of 30~K to 293~K, laying a data foundation for the device selection and technical improvement of subsequent low-temperature detectors. The test results show that all three types of SiPMs can work normally in a low-temperature environment, but there are obvious differences in the performance of products from different manufacturers. 

\section{Experimental setup}\label{sec2}
\subsection{Cryogenic system}\label{subsec1}
\begin{figure}[h]
\centering
\includegraphics[width=0.4\textwidth]{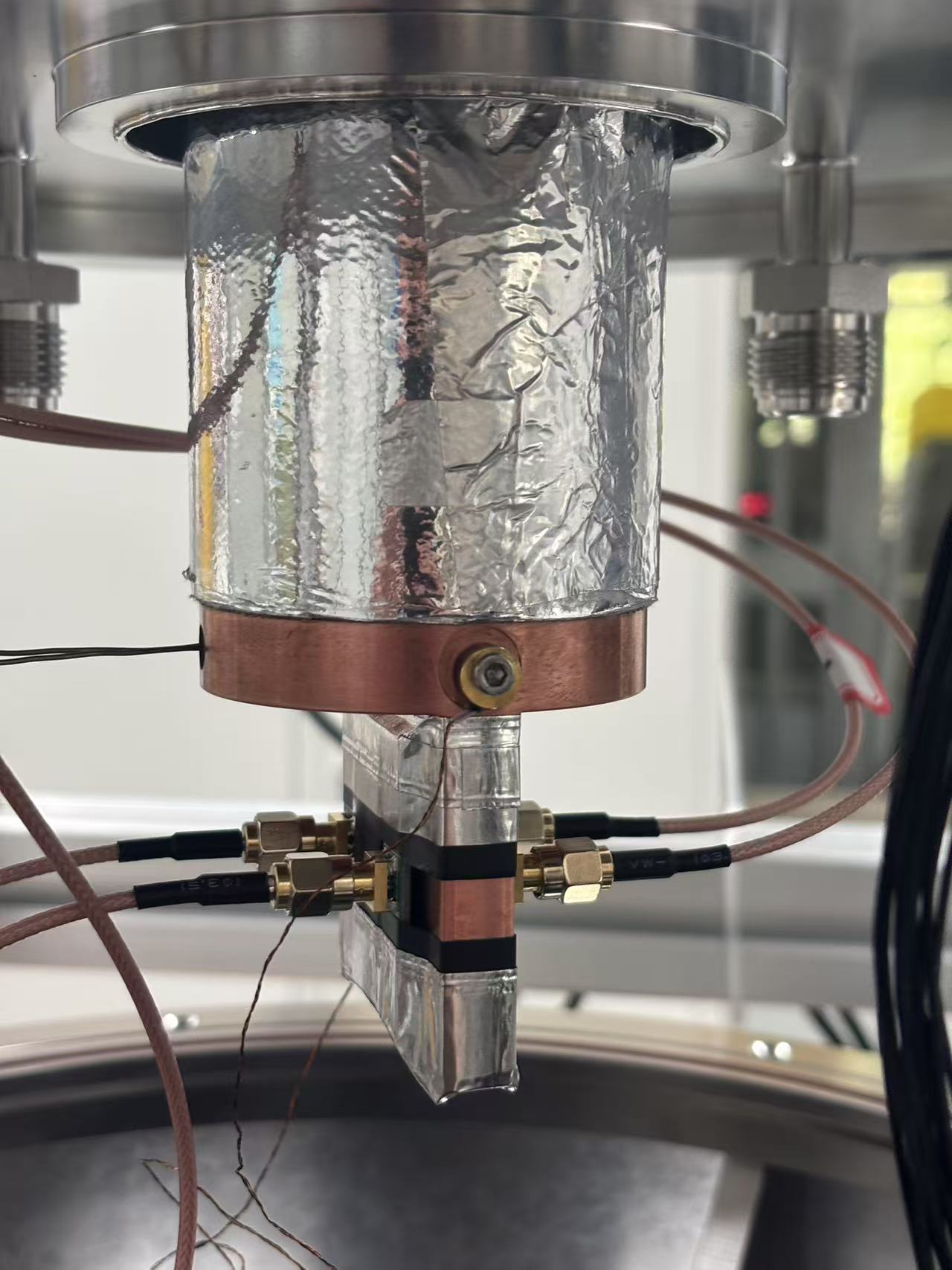}
\qquad
\includegraphics[width=0.4\textwidth]{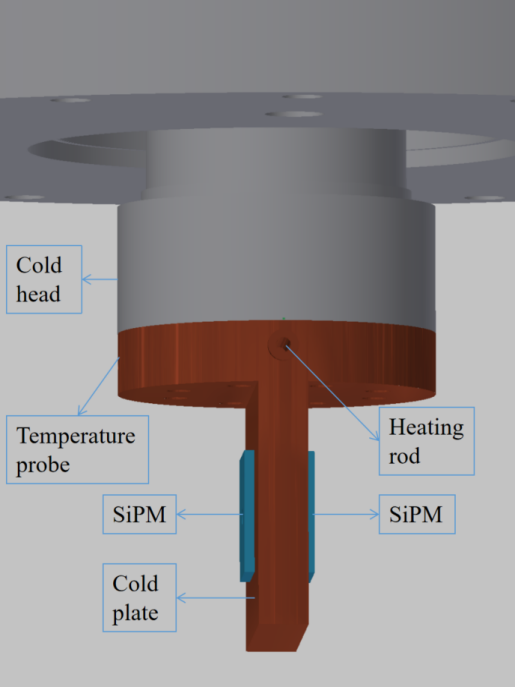}
\caption{The left figure is a physical diagram. The right figure annotates the placement methods of the cold head, cold plate, and SiPM. Apiezon N-type vacuum grease is applied between the SiPM and the cold plate to ensure more uniform heat distribution. SiPMs of the same model are symmetrically installed on both sides of the cold plate. Through dual-channel data comparison, the accuracy of experimental results is improved, and single-point measurement errors are reduced.}\label{fig:1}
\end{figure}

This study developed a set of low-temperature equipment to address the issue that existing devices cannot directly meet the requirements for researching the low-temperature performance of SiPMs. The system employs a cryostat with a vacuum insulation structure (fig.~\ref{fig:1}). Through optimized design based on heat leakage simulation calculations, insulation materials are partially used to reduce heat loss. The coupling structure between the cold plate and the SiPM ensures uniform and efficient cooling of the detector within the temperature range of 30~K to 293~K. The system integrates a Sumitomo CH-104 refrigerator and a 100~W precision heating rod, and uses two PT-100 temperature probes for measurement, thus constructing a reliable basic platform for temperature control.

The temperature control system adopts a Lake Shore 336 temperature controller as the core control unit. It dynamically adjusts the heating power through a high-precision PID algorithm (a control algorithm that combines proportional, integral, and differential links), achieving a temperature stability of \SI{\pm 1} K. The system is designed with a low-heat-load cable lead-out scheme, through which signal lines and cables are led out via vacuum electrode feedthrough. This ensures the stability of the low-temperature environment while guaranteeing the reliability of signal transmission. The successful development of the system not only solves the current experimental needs but also lays a solid foundation for subsequent research on low-temperature detectors.

\begin{figure}[h]
\centering
\includegraphics[width=0.8\textwidth]{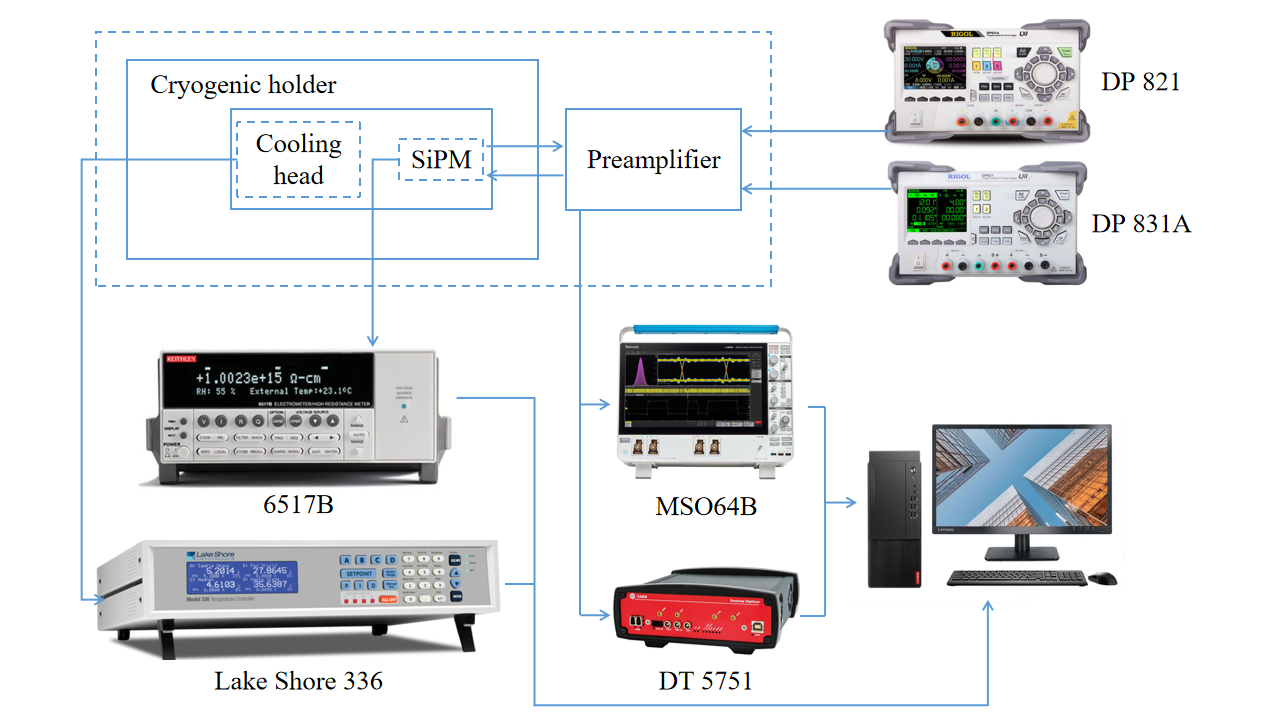}
\caption{Schematic diagram of SiPM low-temperature readout.}\label{fig:2}
\end{figure}

\subsection{Test system}\label{subsec2}
\begin{figure}[h]
\centering
\begin{tabular}{@{}cc@{}}
\includegraphics[width=0.4\textwidth,height=5cm,keepaspectratio]{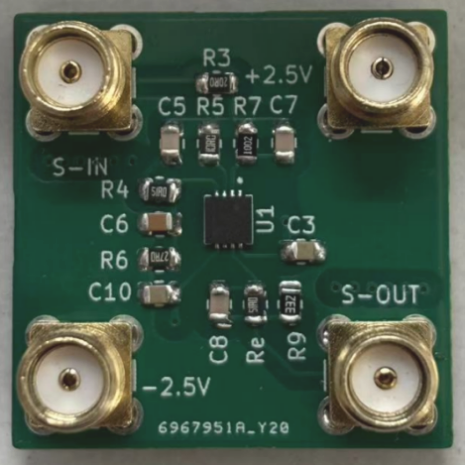} & 
\includegraphics[width=0.4\textwidth,height=5cm,keepaspectratio]{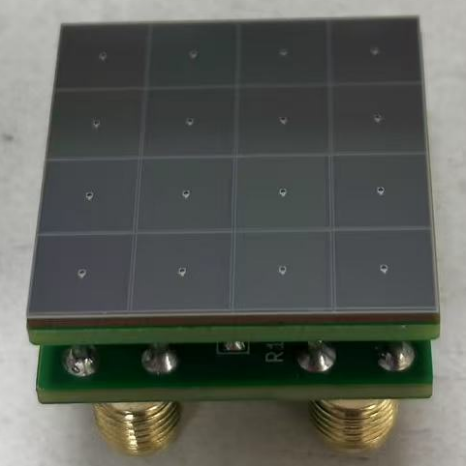} \\
\includegraphics[width=0.4\textwidth,height=5cm,keepaspectratio]{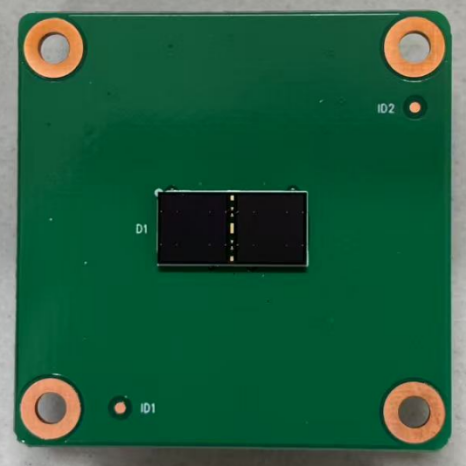} & 
\includegraphics[width=0.4\textwidth,height=5cm,keepaspectratio]{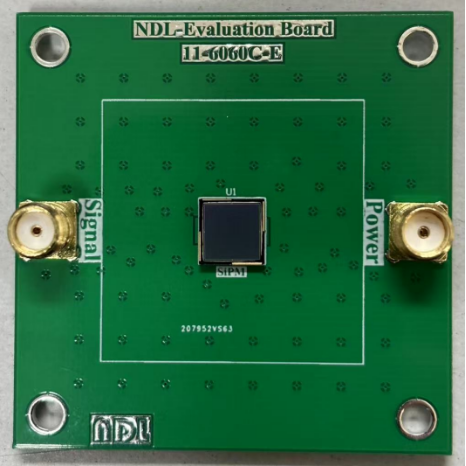}
\end{tabular}
\caption{SiPM and preamplifier. The top-left image shows the preamplifier with a gain of 10×. The top-right image shows the 4 × 4 array SiPM developed by Hamamatsu, with each chip measuring 6 mm × 6 mm. The bottom-left image shows the 2 × 1 dual-chip SiPM developed by Broadcom, with each chip measuring 6.14 mm × 6.14 mm. The bottom-right image shows the monolithic SiPM developed by NDL, with dimensions of 6.24 mm × 6.24~mm.}\label{fig:3}
\end{figure}

The experiment established a comprehensive cryogenic test platform (fig.~\ref{fig:2}) to investigate the performance characteristics of  SiPMs under low-temperature conditions. The system employs a vacuum-insulated cryogenic chamber, where the SiPM samples under test (fig.~\ref{fig:3}, including Hamamatsu S14161-6050HS-04, Broadcom AFBR-S4N66P024M 2 × 1 array, and NDL EOR20 11-6060-E-P models) are mounted on the cold plate. By setting distinct temperature set points, the platform systematically investigates the temperature-dependent variations of their key operational parameters.

The test system is equipped with two independent power supplies: a RIGOL DP821 unit provides adjustable bias voltage (up to 60~V) for the SiPMs, while a RIGOL DP831A supplies \SI{\pm 2.5} V operating voltage to the preamplifier. For signal acquisition, the SiPM output signals are amplified 10-fold by the preamplifier and recorded via an oscilloscope, with simultaneous dark count measurements performed using a DT5751 data acquisition module. The system also integrates a Keithley 6517B electrometer to characterize the I-V curves of the SiPMs. The measured breakdown voltages are cross-verified against values derived from gain calculations, ensuring the accuracy of experimental data.

\section{Results and discussion}\label{sec3}
\begin{figure}[h]
\centering
\includegraphics[width=0.5\textwidth]{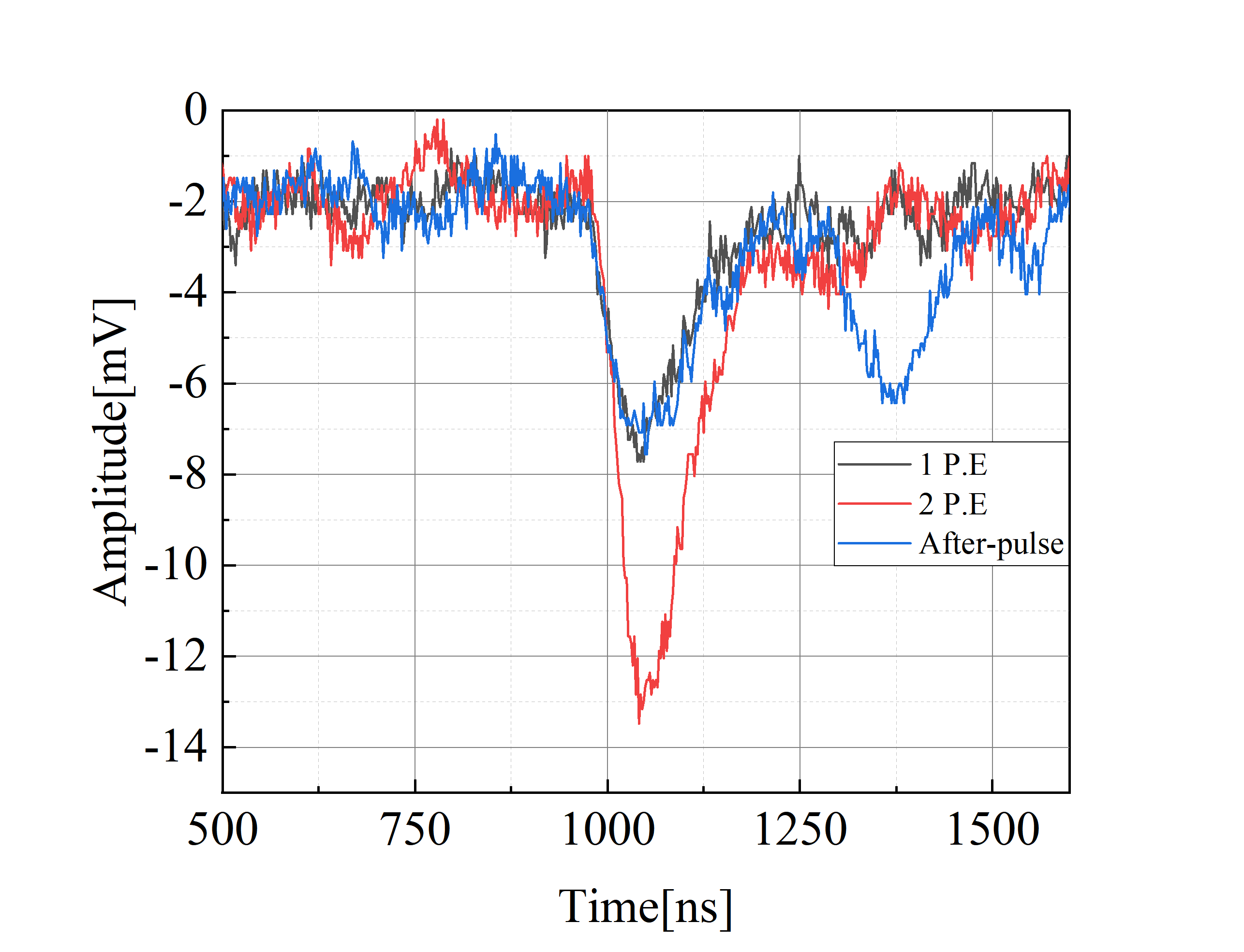}
\caption{The figure above shows the single-photon, two-photon, and after-pulse waveforms of Hamamatsu SiPM at 40~K and a voltage of 36.5~V.}\label{fig:4}
\end{figure}

\begin{figure}[h]
\centering
\includegraphics[width=0.6\textwidth]{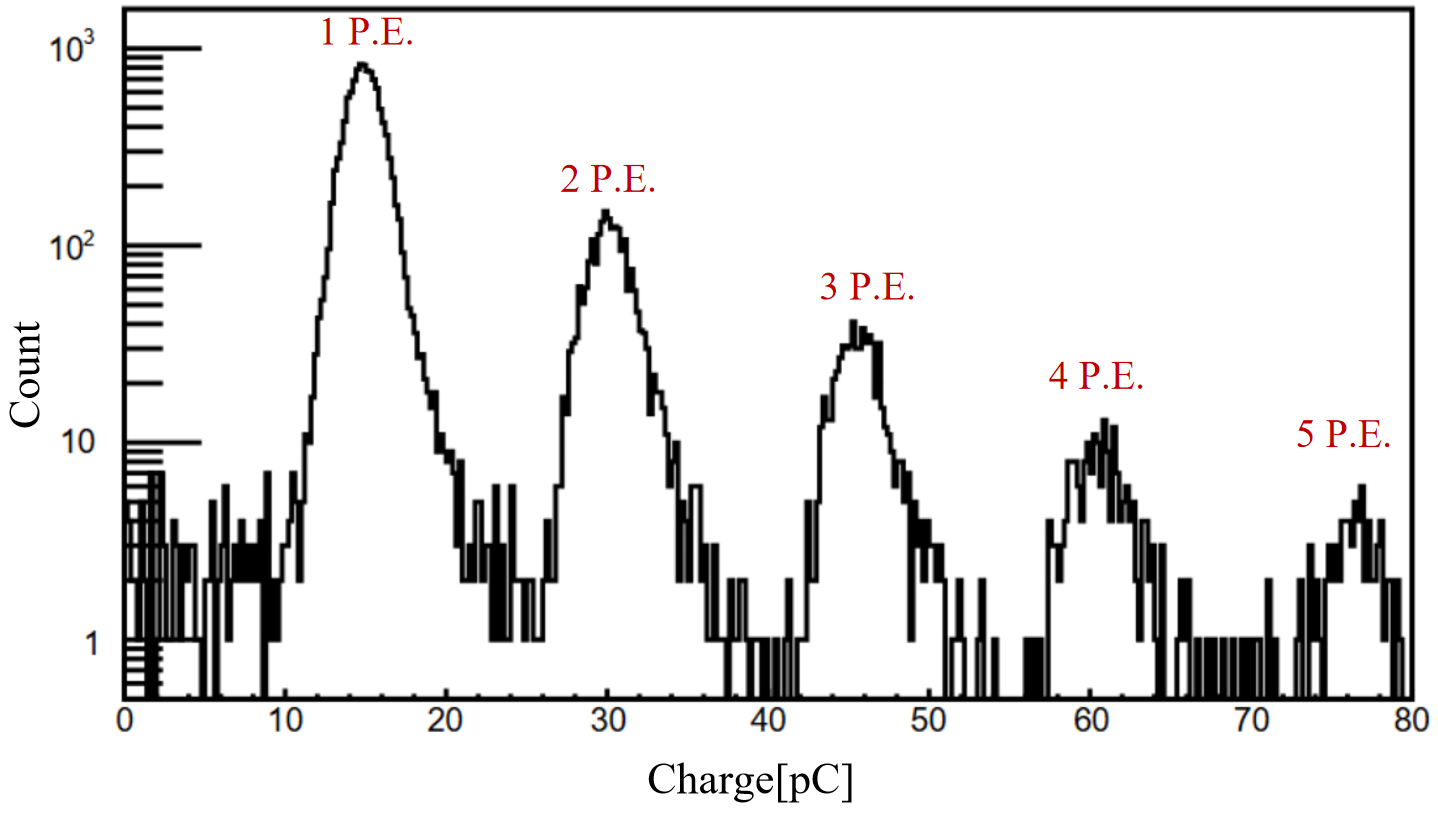}
\caption{The figure above shows the energy spectrum of Hamamatsu SiPM at 40~K and a voltage of 36.5~V.}\label{fig:5}
\end{figure}

Based on raw waveform data acquired from the oscilloscope, this study developed a data processing framework using the CERN ROOT software \cite{24}. By performing peak detection and signal interval integration on each individual waveform (fig.~\ref{fig:4}), the program generates photoelectron spectra of the SiPMs, clearly resolving single- and multi-photoelectron peaks (fig.~\ref{fig:5}). Such characteristic spectra not only provide an intuitive assessment of SiPM operational performance but also enable a quantitative evaluation of their background noise characteristics.

\subsection{Gain}\label{subsec1}
The gain of a SiPM characterizes its single-photon signal amplification capability, which fundamentally originates from the avalanche multiplication effect in avalanche photodiode (APD) microcells. When the bias voltage exceeds the breakdown voltage, photon-generated electron-hole pairs undergo impact ionization under a high electric field (10\(^5\)~V/cm), triggering an avalanche chain reaction. This process ultimately produces a large number of secondary charge carriers within the microcell, with the total count corresponding to the gain. The calculation formula is expressed as follows:

\begin{equation}
\text{Gain} = \frac{C_{\text{pixel}} \times (V_{\text{bias}} - V_{\text{bd}})}{e} = \frac{Q}{A \times e}
\end{equation}

Let \(C_{pixel}\) denote the microcell junction capacitance, \(V_{bias}\) the bias voltage, \(V_{bd}\) the breakdown voltage, \( e \) the elementary charge, \( Q \) the charge per photoelectron, and \( A \) the amplification factor of the preamplifier. 

\begin{figure}[h]
\centering
\includegraphics[width=.495\textwidth]{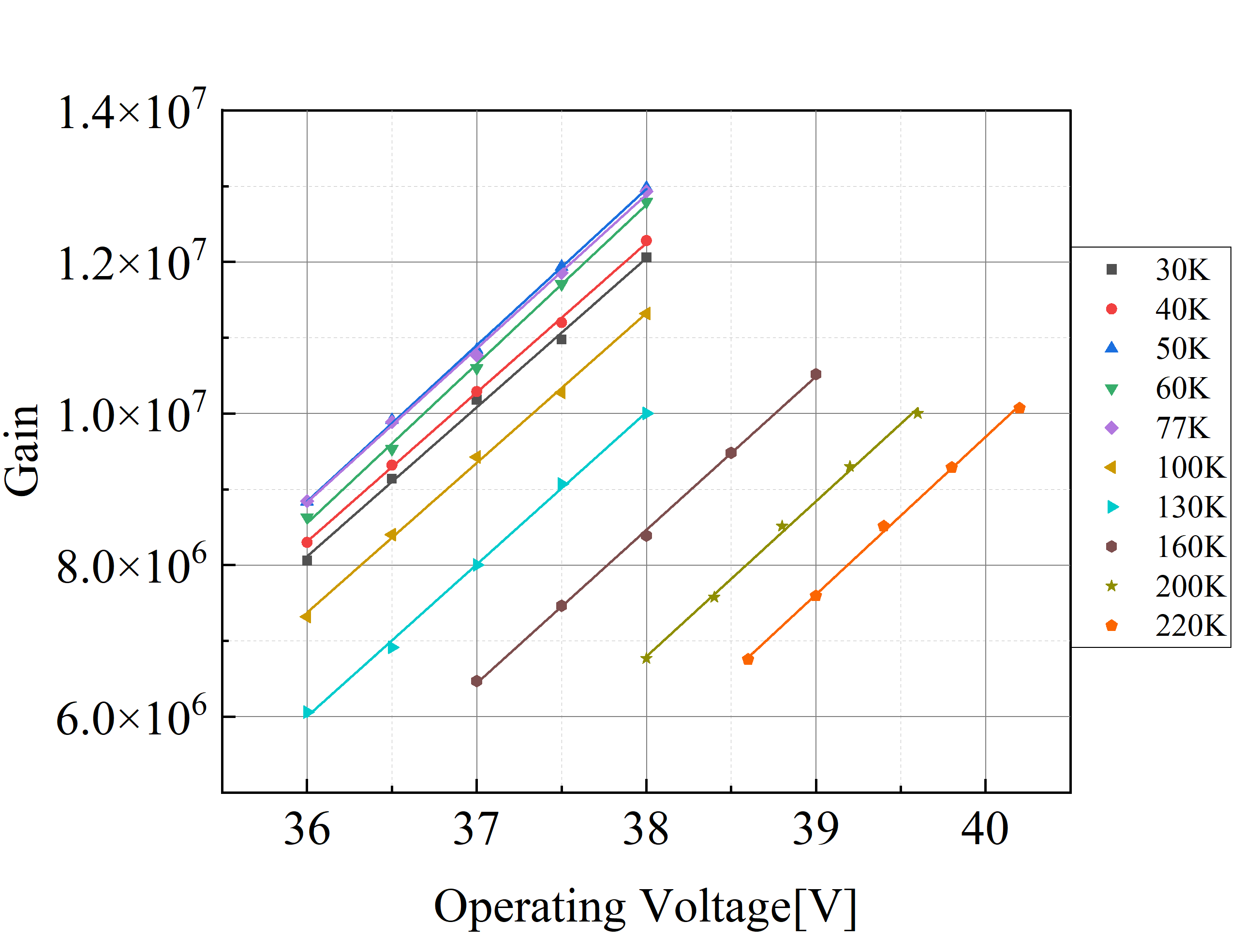}
\includegraphics[width=.495\textwidth]{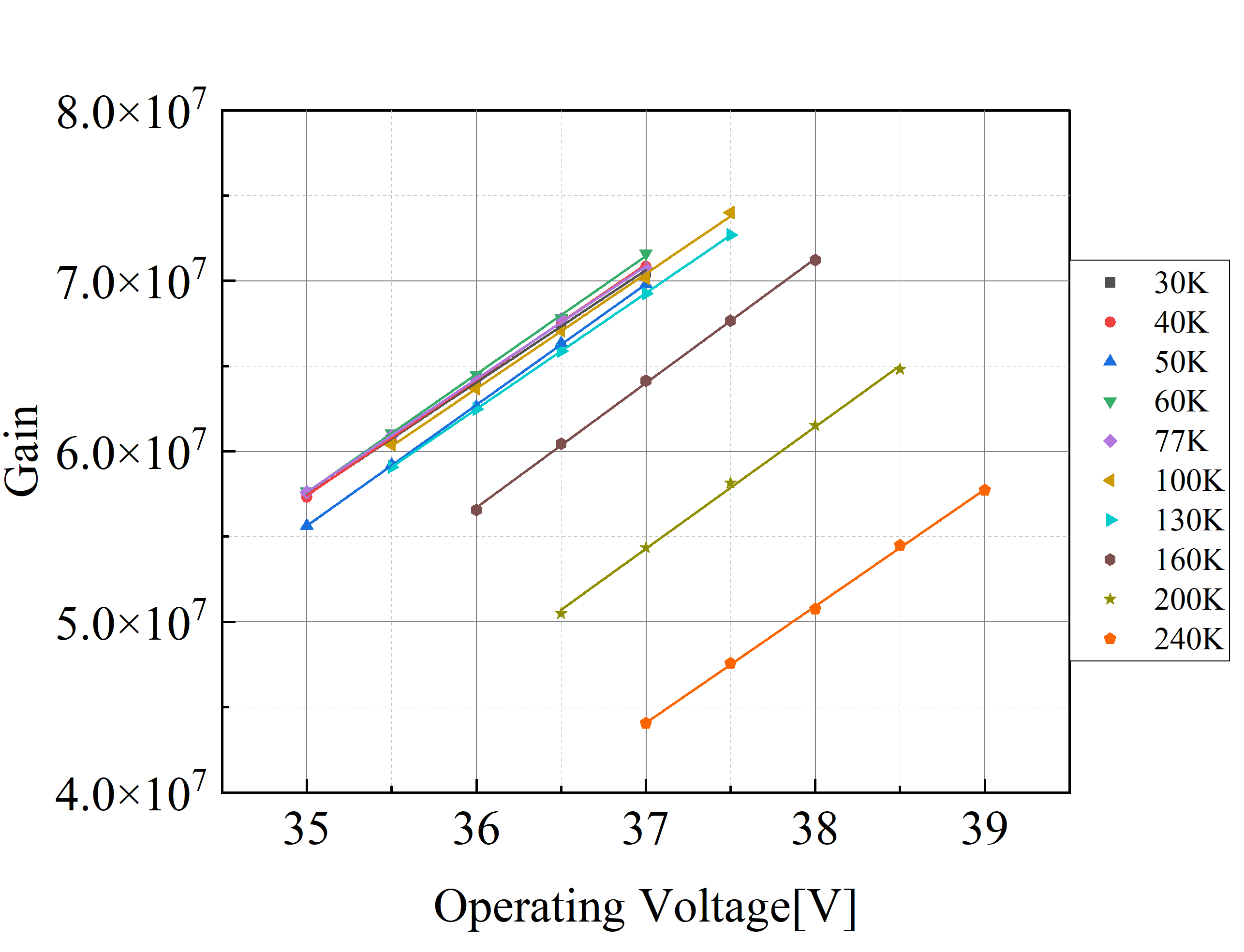}
\includegraphics[width=.495\textwidth]{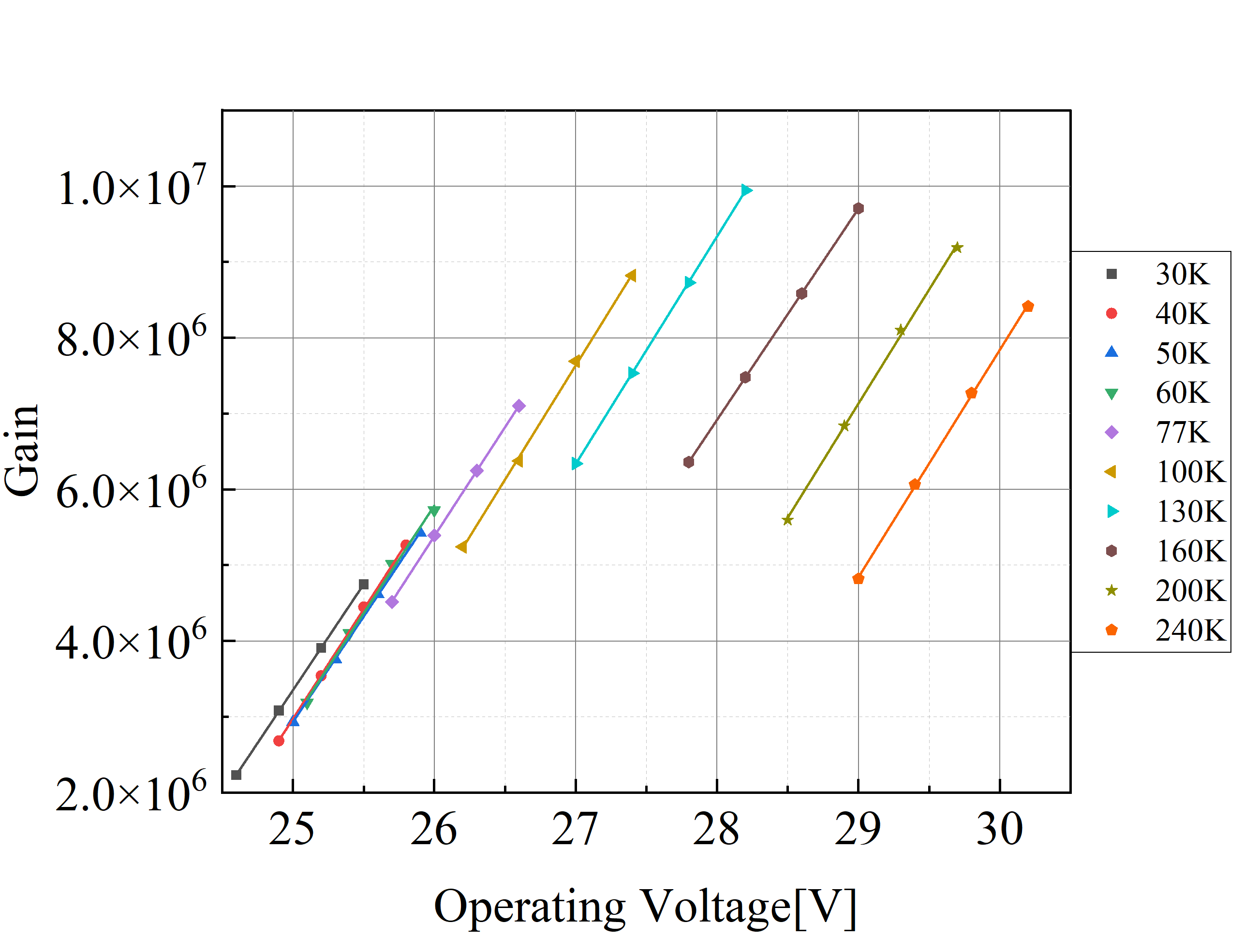}
\includegraphics[width=.495\textwidth]{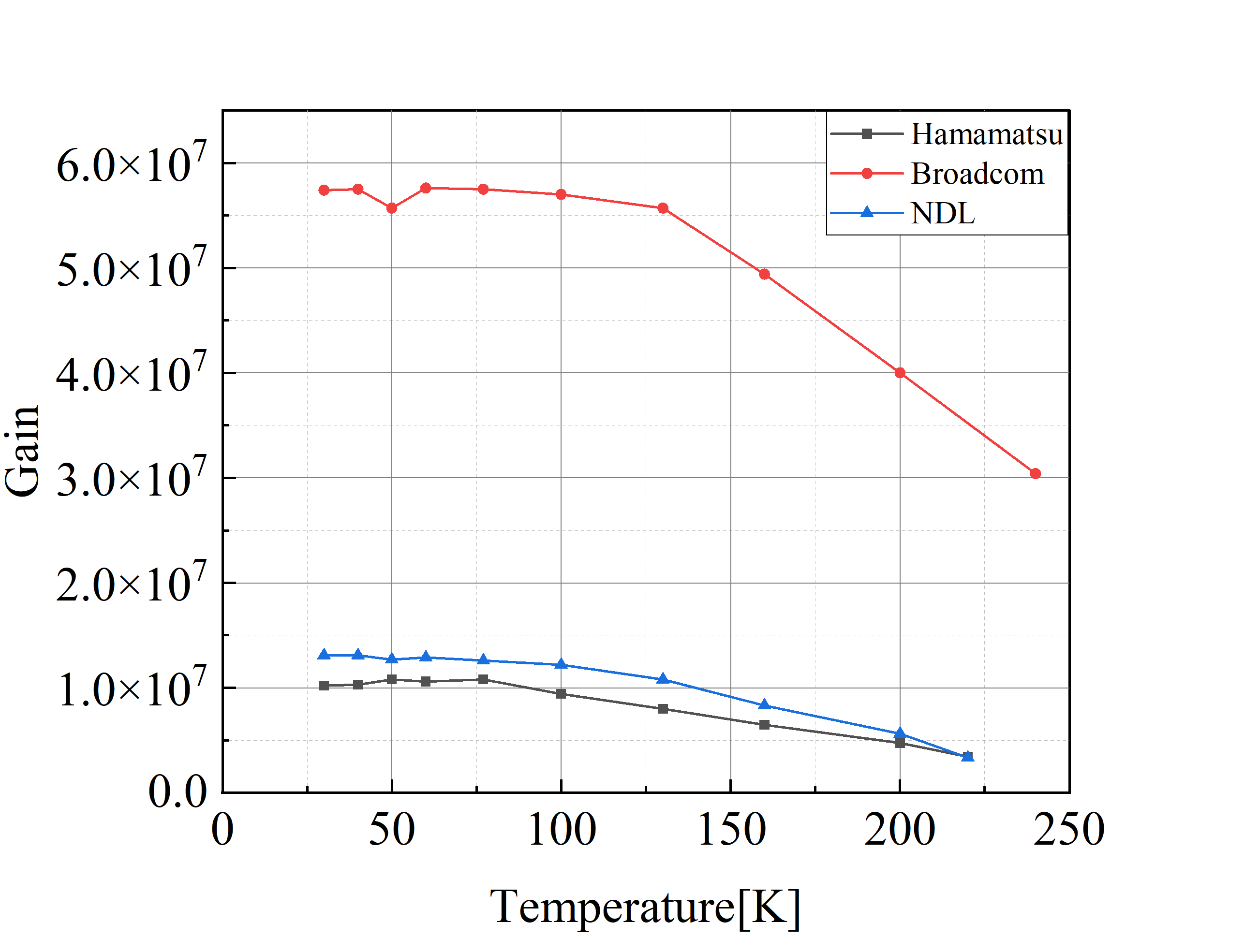}
\caption{The upper left graph shows the relationship between gain and operating voltage of Hamamatsu SiPM under the temperature condition from 30~K to 220~K. The upper right graph shows the relationship between gain and operating voltage of Broadcom SiPM under the temperature condition from 30~K to 240~K. The lower left graph shows the relationship between gain and operating voltage of NDL SiPM under the temperature condition from 30~K to 240~K. The lower right graph shows the gain variation with temperature of Hamamatsu SiPM at 37~V, Broadcom SiPM at 35~V, and NDL SiPM at 28.5~V.}\label{fig:6}
\end{figure}

The operating environment temperature of the SiPM was precisely regulated by a temperature controller, and the relationship between the gain of the SiPM and the operating voltage under different temperature conditions was systematically investigated (fig.~\ref{fig:6}). The experimental results show that the gain of the SiPM has an obvious linear relationship with the operating voltage at the same temperature. Within a certain temperature range, the gain of the SiPM increases as the ambient temperature decreases. Since the light yield of the crystal adopts a relative measurement method, to maintain the consistency of signal readout, the impedance of the SiPM is configured as 50 \(\Omega\) by default. Through the linear fitting analysis of the gain, it is not only possible to calculate the breakdown voltage of the SiPM at different temperatures but also to deeply reveal the influence law of temperature changes on the gain characteristics of the SiPM.

\subsection{Breakdown voltage}\label{subsec2}
The breakdown voltage (\(V_{bd}\)) of an SiPM is a critical parameter determining its operational performance. This study employed two complementary methods to determine \(V_{bd}\): on the one hand, by identifying the inflection point voltage in the I-V characteristic curve where the current exhibits a steep surge. On the other hand, by measuring gain characteristics under different bias voltages and linearly extrapolating to the voltage value at zero gain. The mutual validation of results from both methods enhances the reliability of \(V_{bd}\) measurements.

\begin{figure}[h]
\centering
\includegraphics[width=.495\textwidth]{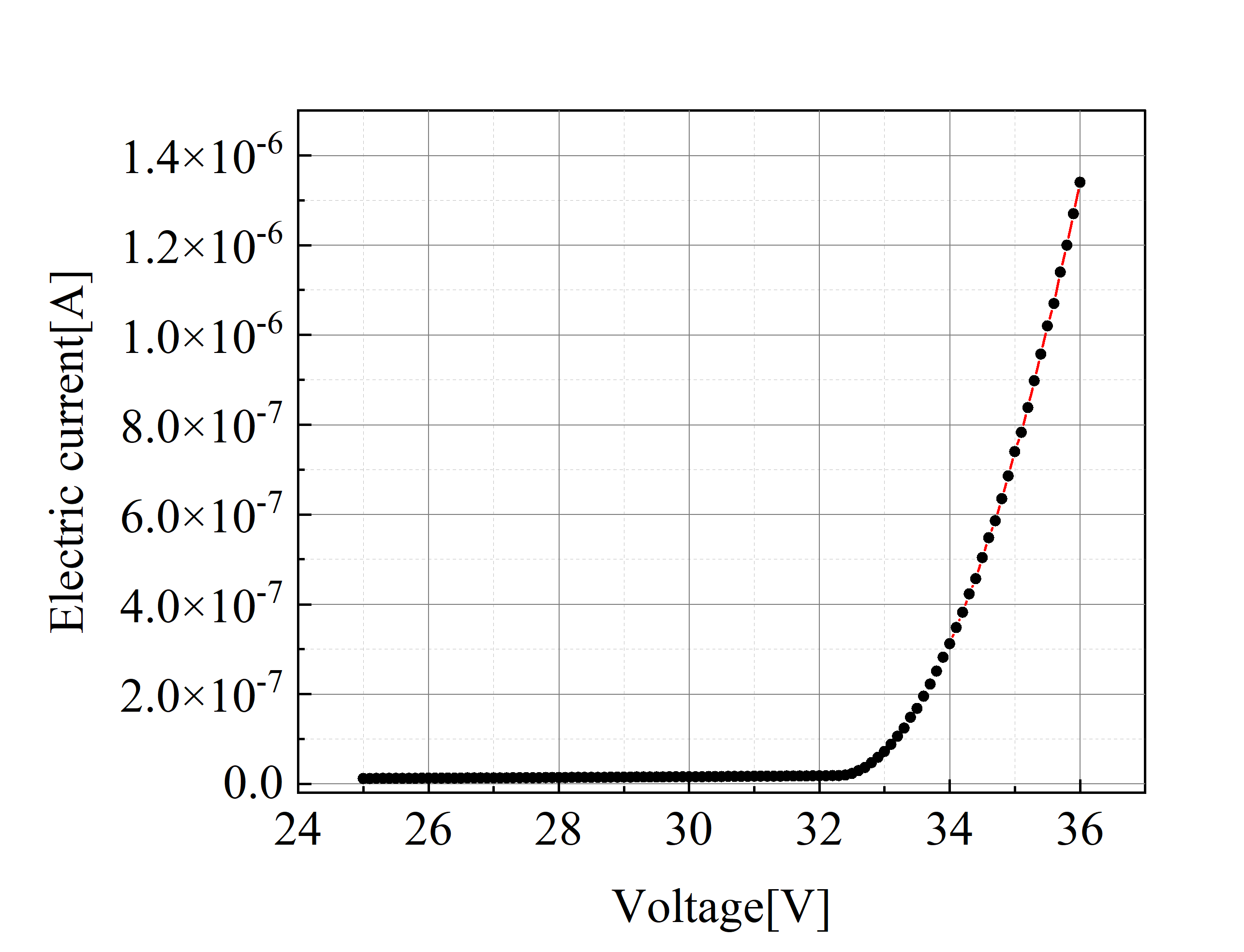}
\includegraphics[width=.495\textwidth]{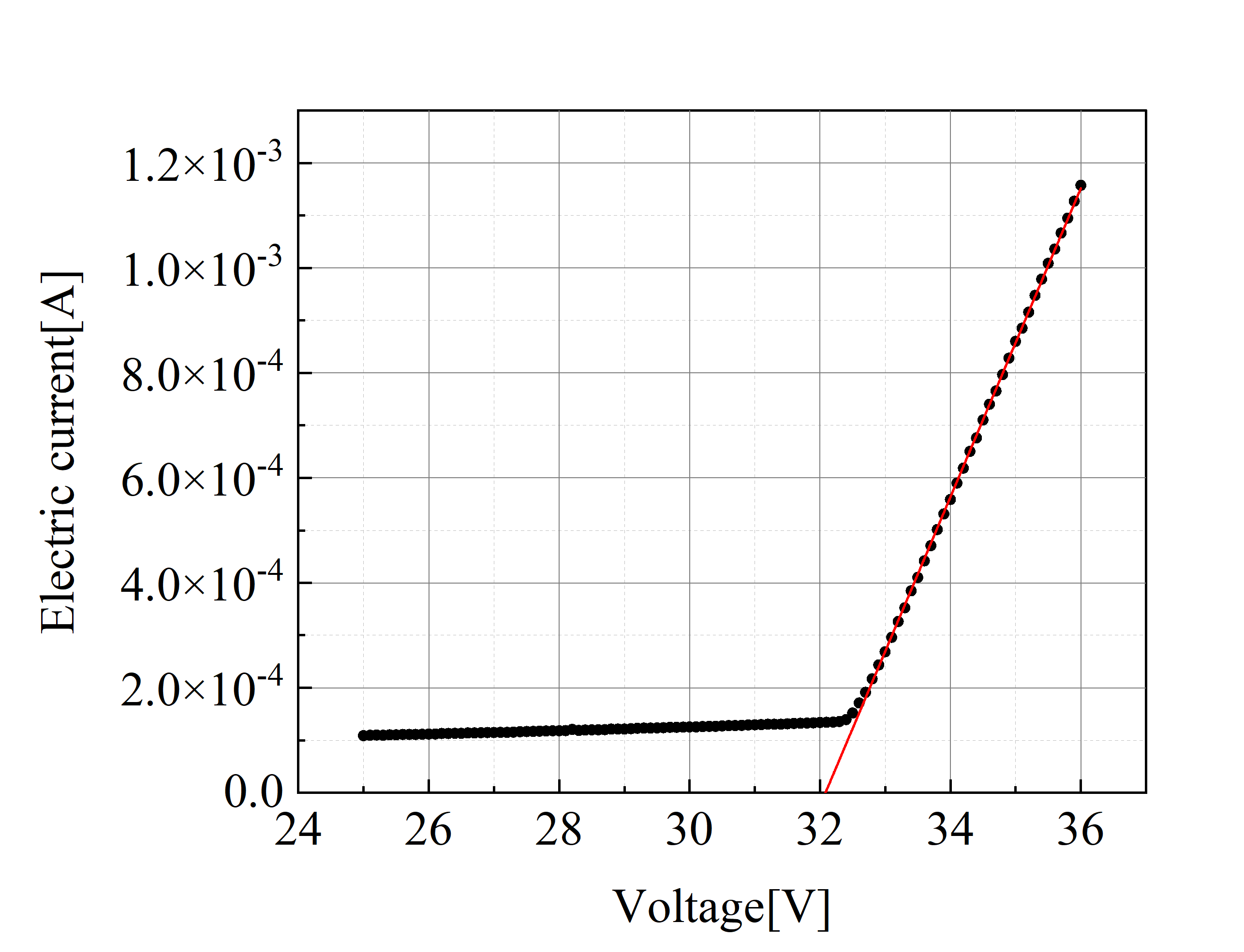}
\caption{The left figure shows the I-V curve of the Broadcom SiPM at 293~K, and the right figure shows the $\sqrt{I}$-V curve of the Broadcom SiPM at 293~K. The \(V_{bd}\) of the SiPM is obtained by fitting the $\sqrt{I}$-V curve, which is 32.1~V, basically consistent with the \(V_{bd}\) tested by the manufacturer.}\label{fig:7}
\end{figure}

The \(V_{bd}\) is determined through I-V curve testing (fig.~\ref{fig:7}): By applying a reverse bias voltage to the SiPM and precisely measuring its dark current characteristics. 

During the testing process, as the bias voltage approaches the \(V_{bd}\), the device current exhibits an exponential increase due to the triggering of the avalanche effect. The inflection point voltage in this curve is identified as the \(V_{bd}\). A high-precision electrometer, Keithley 6517B, is employed for scanning with a step size of 0.1~V, while maintaining a stable test environment temperature (\SI{\pm 1}{~K}). At room temperature, the current increases from $10^{-8}$ A in the non-breakdown state to $10^{-6}$ A in the breakdown state. Due to the fact that the step size is set within a certain range, accurate measurement of the \(V_{bd}\) cannot be achieved. The precise value of the \(V_{bd}\) can be obtained through the fitting of the $\sqrt{I}$-V curve \cite{25}, and the intercept of the linear fitting on the horizontal axis is the \(V_{bd}\).

\begin{figure}[h]
\centering
\includegraphics[width=0.50\textwidth]{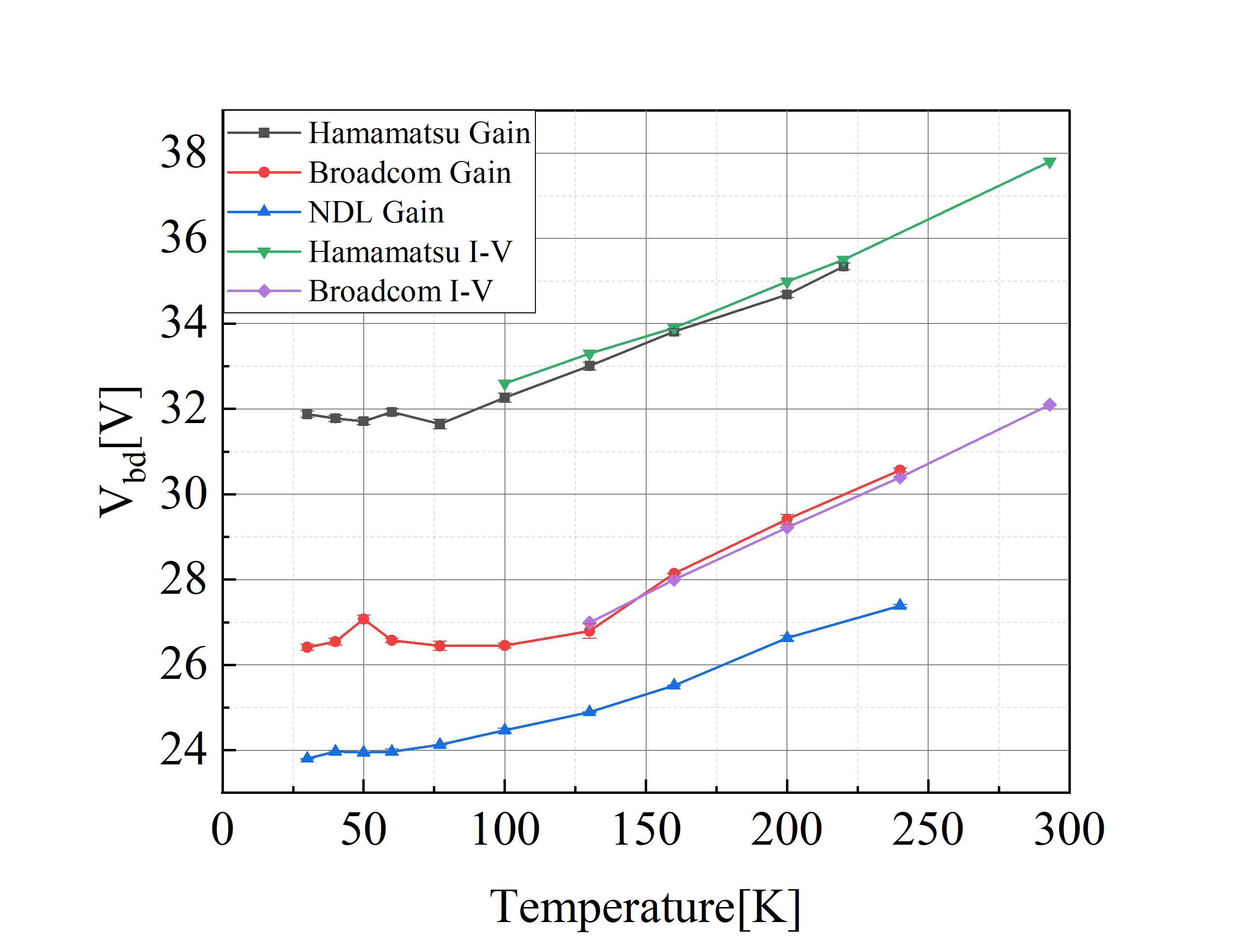}
\caption{
The \(V_{bd}\) is determined through gain fitting, with the voltage value corresponding to zero gain taken as the \(V_{bd}\). The \(V_{bd}\) of the Hamamatsu SiPM varies with temperature in the range of 30~K to 220~K, while that of the Broadcom SiPM and NDL SiPM changes with temperature between 30~K and 240~K. Due to the 4 × 4 array structure of the Hamamatsu SiPM, which results in a relatively large area, the photon pile-up effect is significant at 240~K, leading to poor waveform integration results. The errors displayed on the Y-axis originate from the fitting of the gain. The \(V_{bd}\) was also tested via I-V curve measurements, examining the temperature-dependent characteristics of the \(V_{bd}\) for the Hamamatsu SiPM in the temperature range of 100~K to 293~K and for the Broadcom SiPM in the range of 130~K to 293~K.}\label{fig:8}
\end{figure}

Through gain fitting (fig.~\ref{fig:8}): The \(V_{bd}\) variation was experimentally measured within the temperature range of 30~K to 293~K. The results indicate that for Hamamatsu SiPM and NDL SiPM, the \(V_{bd}\) exhibits a nearly linear relationship with temperature above 77~K, with temperature coefficients of 25.8~mV/K and 20.2~mV/K, respectively. In contrast, Broadcom SiPM demonstrates a linear characteristic only when the temperature exceeds 130~K, with a temperature coefficient of 34.3~mV/K. In the temperature range from 77~K to 293~K, the \(V_{bd}\) decreases as the temperature decreases. This is mainly because the vibration of semiconductor lattice atoms decreases with the decrease of temperature, which leads to the widening of the potential barrier layer in the P-N junction and the increase of the average free path of carriers \cite{26}. Thus, avalanche breakdown can occur with a lower electric field.

During the testing process, no external light source was employed, and the measurements relied solely on the intrinsic dark noise of the SiPM. The research revealed that when the temperature is excessively low, the dark current decreases significantly, making it impossible to accurately determine the \(V_{bd}\) through the I-V curve. Furthermore, due to the adoption of an epitaxial resistance structure in the NDL SiPM, its \(V_{bd}\) cannot be measured using the I-V curve testing method. Within a certain temperature range where both test methods are applicable, the measured \(V_{bd}\) results are basically consistent.

When the temperature drops below 77~K, the \(V_{bd}\) of SiPM deviates from the linear variation trend and even shows an increasing phenomenon, which is mainly attributed to the changes in the low-temperature characteristics of semiconductor materials \cite{26}. The carrier concentration becomes extremely low, requiring a stronger electric field to trigger the avalanche. Meanwhile, the weakening of carrier-phonon scattering will enhance the avalanche efficiency. These effects collectively lead to a special temperature-dependent characteristic of the \(V_{bd}\) in the range of 30~K to 77~K, and even a slight upward abnormal phenomenon is observed.

\subsection{Dark counts}\label{subsec3}
The dark count rate (DCR) of SiPMs, as a critical performance metric, directly reflects the background noise characteristics of the device under dark conditions \cite{27}. Its physical essence lies in the probability of thermally excited carriers triggering the avalanche process. 

\begin{figure}[h]
\centering
\includegraphics[width=0.495\textwidth]{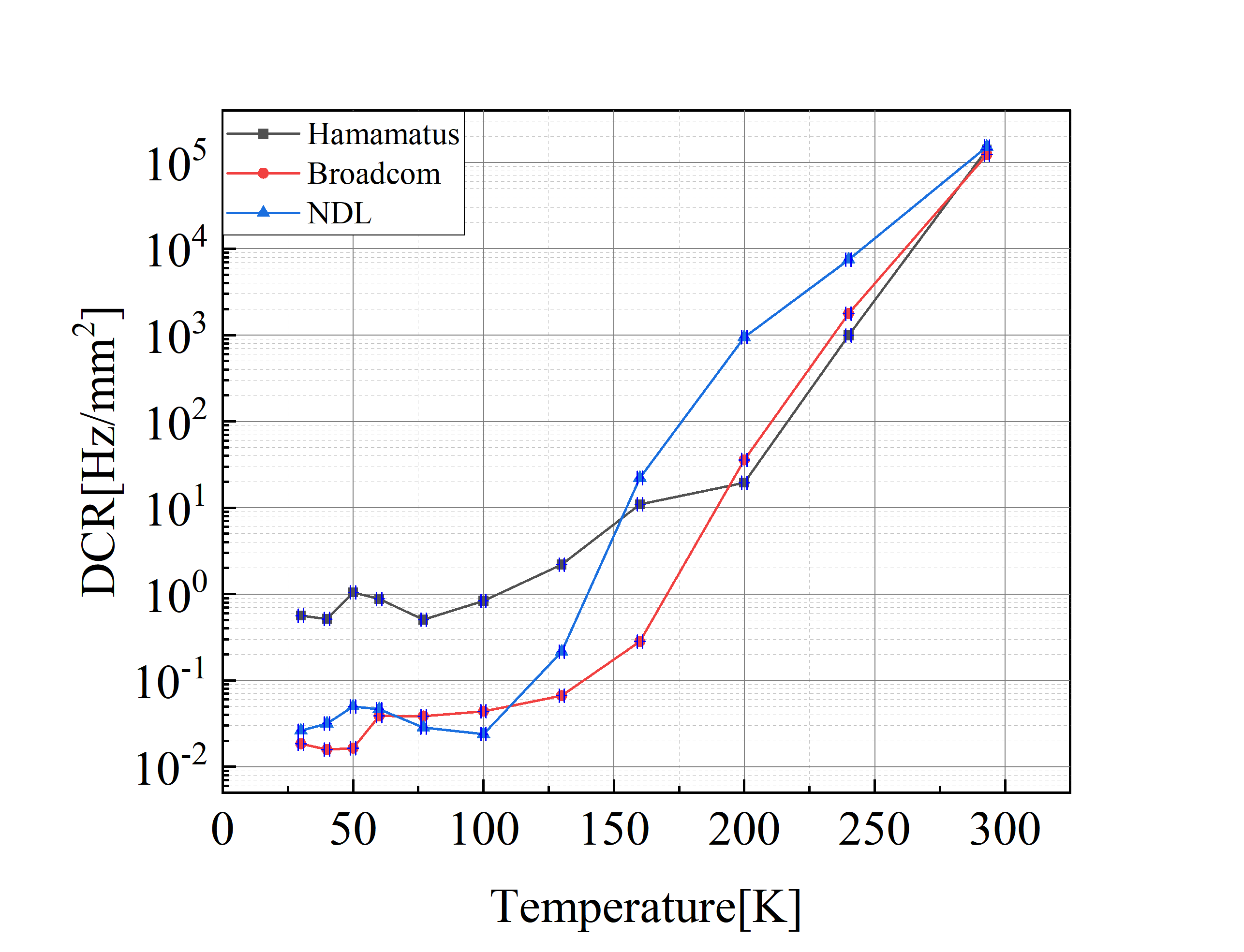}
\includegraphics[width=0.495\textwidth]{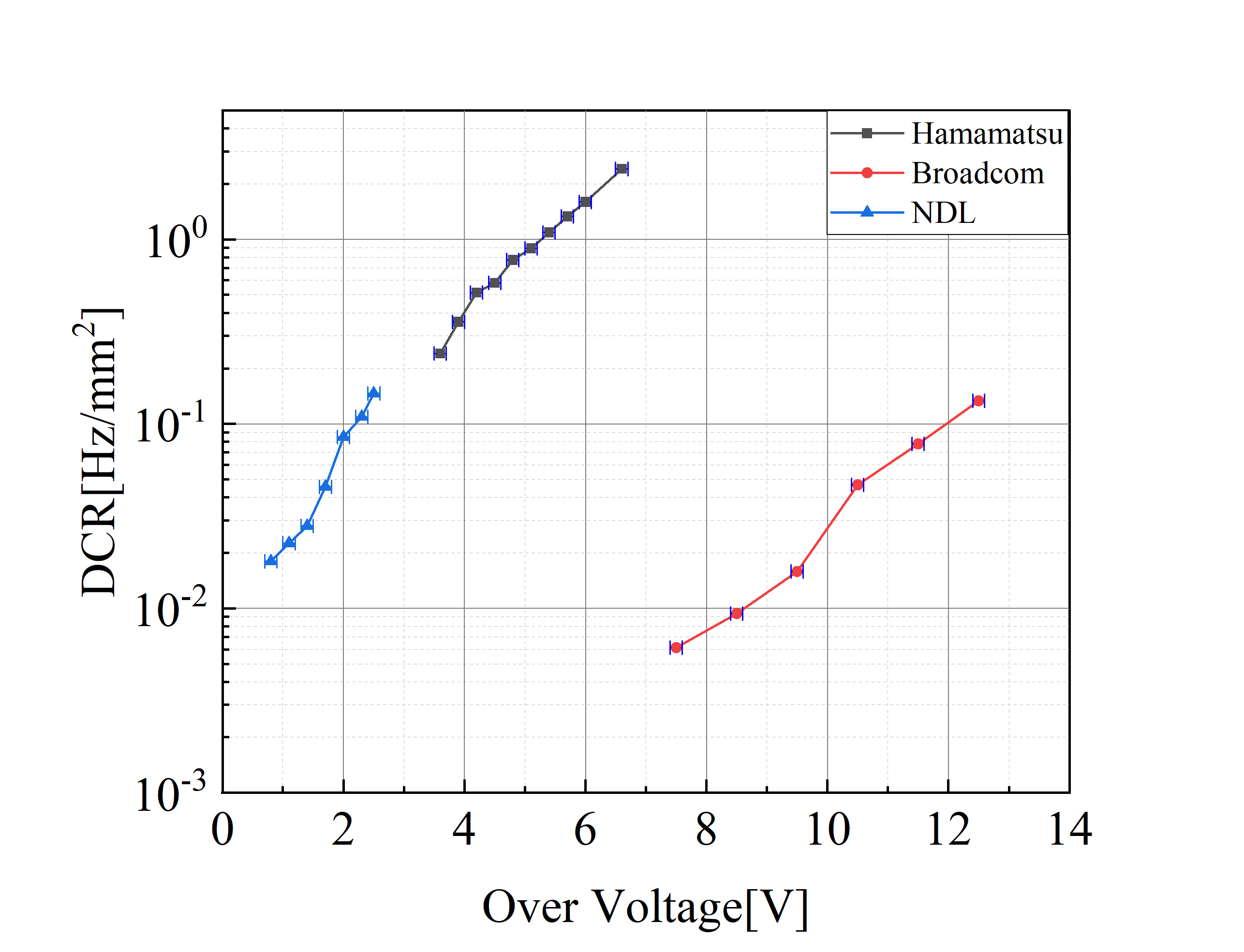}
\caption{The left graph illustrates the variation of DCR with temperature. The overvoltage settings are 4~V for Hamamatsu SiPM, 9~V for Broadcom SiPM, and 2.5~V for NDL SiPM. Different models of SiPMs are set with different bias voltages primarily because each model has a specific optimal operating voltage range. In low-temperature environments, applying an excessively high bias voltage may lead to secondary breakdown in the SiPM. The right panel presents the DCR dependence of the SiPM on operating voltage at 40~K. The error on the X-axis is induced by temperature fluctuations (\SI{\pm 1}{~K}) during testing.}\label{fig:9}
\end{figure}

This study systematically investigates the variation laws of DCR in the dual-parameter space of temperatures from 30~K to 293~K and overvoltage. It is worth noting that when the operating temperature drops from 293~K to 130~K, the DCR sharply decreases from 10\(^5\) \si{\Hz\per\mm^2} to $10^{-1}$ - 10\(^0\) \si{\Hz\per\mm^2}, a reduction of 6 orders of magnitude (fig.~\ref{fig:9}). This result clearly demonstrates the significant effect of low-temperature operation on suppressing thermal noise.

At a fixed temperature of 40~K, increasing the overvoltage of the Hamamatsu SiPM from 3.6~V to 6.6~V causes the DCR to rise from 0.24 \si{\Hz\per\mm^2} to 2.41 \si{\Hz\per\mm^2}. Similarly, increasing the overvoltage of the Broadcom SiPM from 7.5~V to 12.5~V results in a DCR increase from 0.006 \si{\Hz\per\mm^2} to $0.15\,\si{\Hz\per\mm^2}$, while raising the overvoltage of the NDL SiPM from 0.8~V to 2.5~V leads to a DCR increase from 0.02 \si{\Hz\per\mm^2} to 0.15 \si{\Hz\per\mm^2}. This nonlinear growth relationship can be attributed to two factors: on one hand, the enhanced electric field increases the probability of avalanche triggering. On the other hand, excessively high bias voltages may contribute to secondary processes such as tunneling effects.  

By establishing a comprehensive DCR parameter response map, this study not only deepens the understanding of the noise mechanisms in SiPMs but also provides critical experimental insights for device selection and system design. 

\subsection{After-pulse}\label{subsec4}
\begin{figure}[h]
\centering
\includegraphics[width=0.6\textwidth]{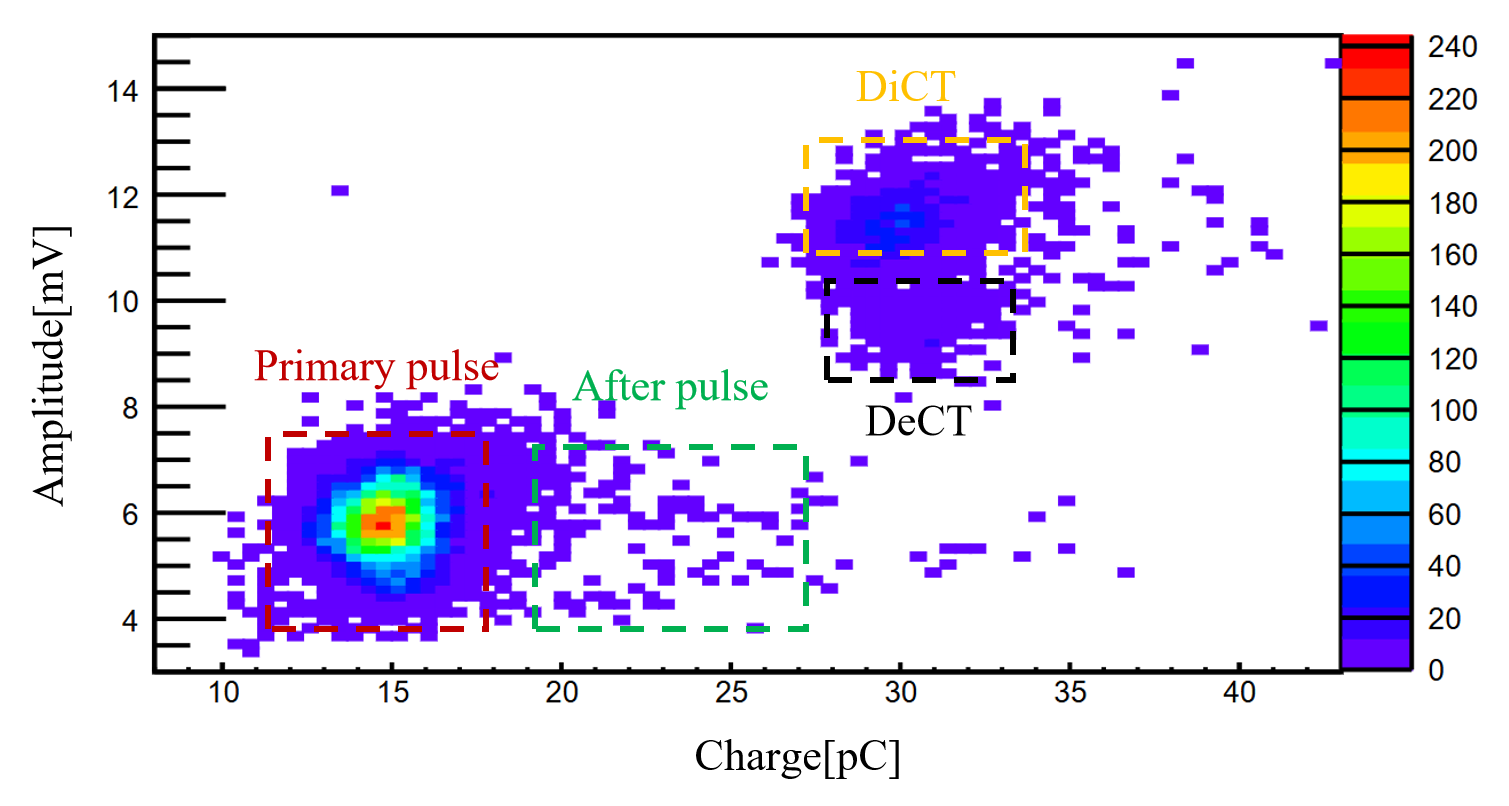}
\caption{Charge-amplitude scatter distribution of SiPM. Herein, the red dashed box corresponds to the single-photon (1 P.E.) signal, the green dashed box contains after-pulse events, the orange dashed box marks direct crosstalk (DiCT) events, and the black dashed box marks delayed crosstalk (DeCT) events. Only the after-pulse generated after single-photon signals are considered. Firstly, in terms of timing, they occur after the initial single-photon avalanche event. Secondly, the amplitude of the after-pulse signal is slightly smaller than that of the single-photon signal. In the scatter plot, due to the integration characteristic of the signal acquisition system, after-pulse events and single-photon signals are recorded simultaneously within the same integration time window. This causes the charge exhibited by after-pulses to be between 1 P.E. and 2 P.E., while their signal amplitude is the same as that of single-photon events.}\label{fig:10}
\end{figure}

The measurement of after-pulse is also based on SPE spectra \cite{27,28,29}. After-pulse arises from secondary pulses formed by the delayed release of charge carriers trapped during the avalanche process. These pulses typically exhibit smaller amplitudes than the original signals (fig.~\ref{fig:4}), manifesting as "small pulses" that are difficult to distinguish from genuine SPE signals, thereby degrading photon-counting resolution. Therefore, it is crucial to study the influence of bias voltage on the after-pulse rate at 40~K, so as to reduce its interference with real signals. The charge-amplitude two-dimensional distribution scatter plot shown below (fig.~\ref{fig:10}) clearly demonstrates the response characteristics of the Hamamatsu SiPM operating at 40~K with a working voltage of 36.5~V.

\begin{figure}[h]
\centering
\includegraphics[width=0.50\textwidth]{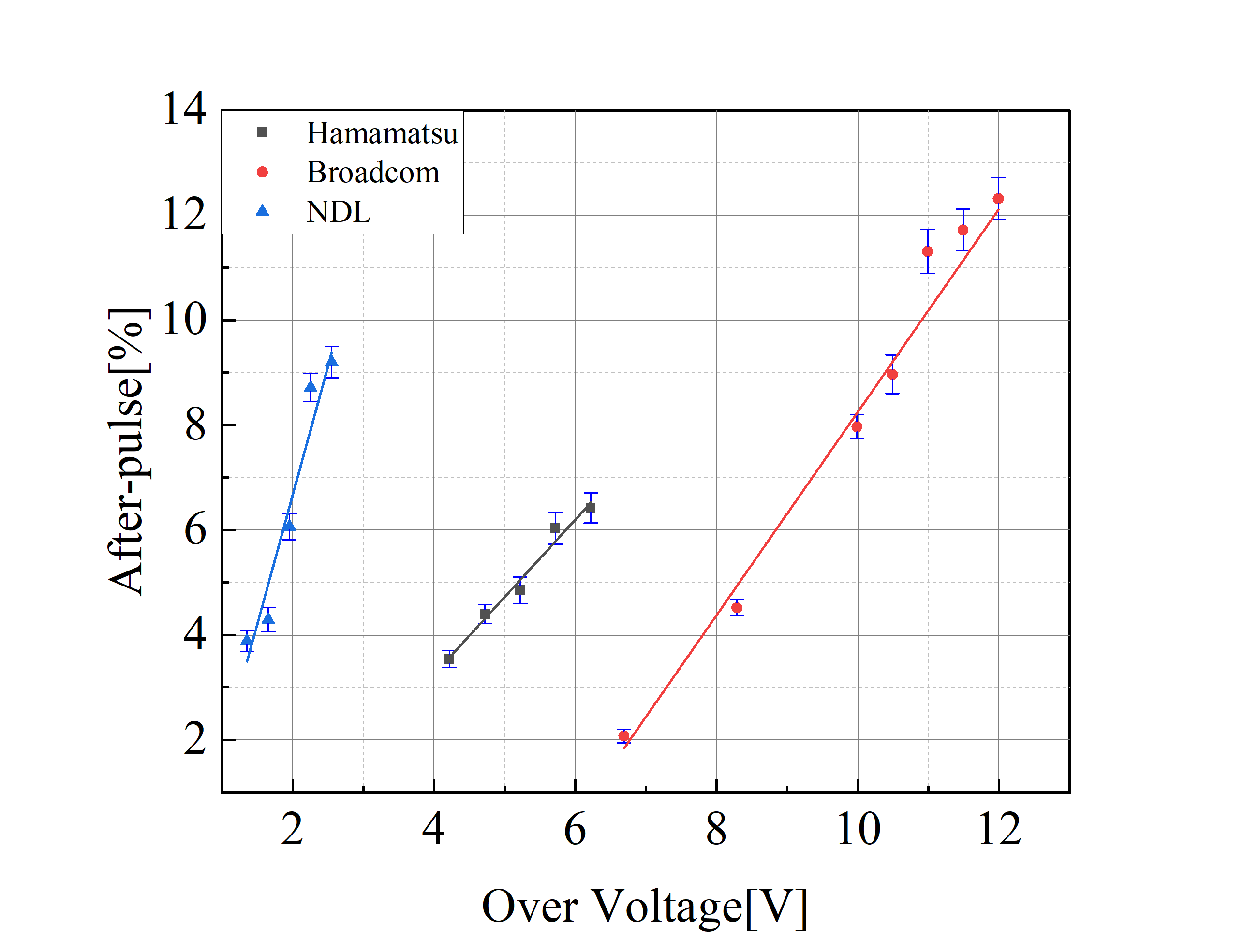}
\caption{The probability of after-pulse for three SiPMs at 40~K as a function of overvoltage. The error bars on the Y-axis represent statistical errors generated when calculating the probability of after-pulse.}\label{fig:11}
\end{figure}

Through analysis of after-pulse probabilities in three SiPM models under varying overvoltage (fig.~\ref{fig:11}), it is observed that within a specific bias voltage range, the after-pulse probability increases with higher overvoltage. This analytical method provides reliable quantitative insights into SiPM performance characteristics. Further improvements in measurement accuracy can be achieved by optimizing parameters such as peak detection algorithms and integration time windows.

\subsection{Internal crosstalk}\label{subsec5}
Internal crosstalk (iCT) \cite{30} is a phenomenon where secondary photons are absorbed by adjacent single-photon avalanche diode (SPAD) and trigger additional avalanches within the same SiPM. Depending on whether electrons are generated in the depletion region, iCT can be further divided into direct crosstalk (DiCT) and delayed crosstalk (DeCT).

DiCT \cite{31} is the secondary signal generated when secondary photons escape to the depletion layer of a nearby second SPAD unit and induce a secondary avalanche. DeCT \cite{32} occurs when secondary photons are absorbed by the substrate in the non-depletion layer of the second SPAD unit. This process can also generate electrons. However, since the electrons are in the non-depletion layer at this time, an avalanche cannot occur. Instead, the electrons will drift toward the depletion layer under the action of an electric field and may trigger an avalanche after entering the depletion layer.

\begin{figure}[h]
\centering
\includegraphics[width=0.50\textwidth]{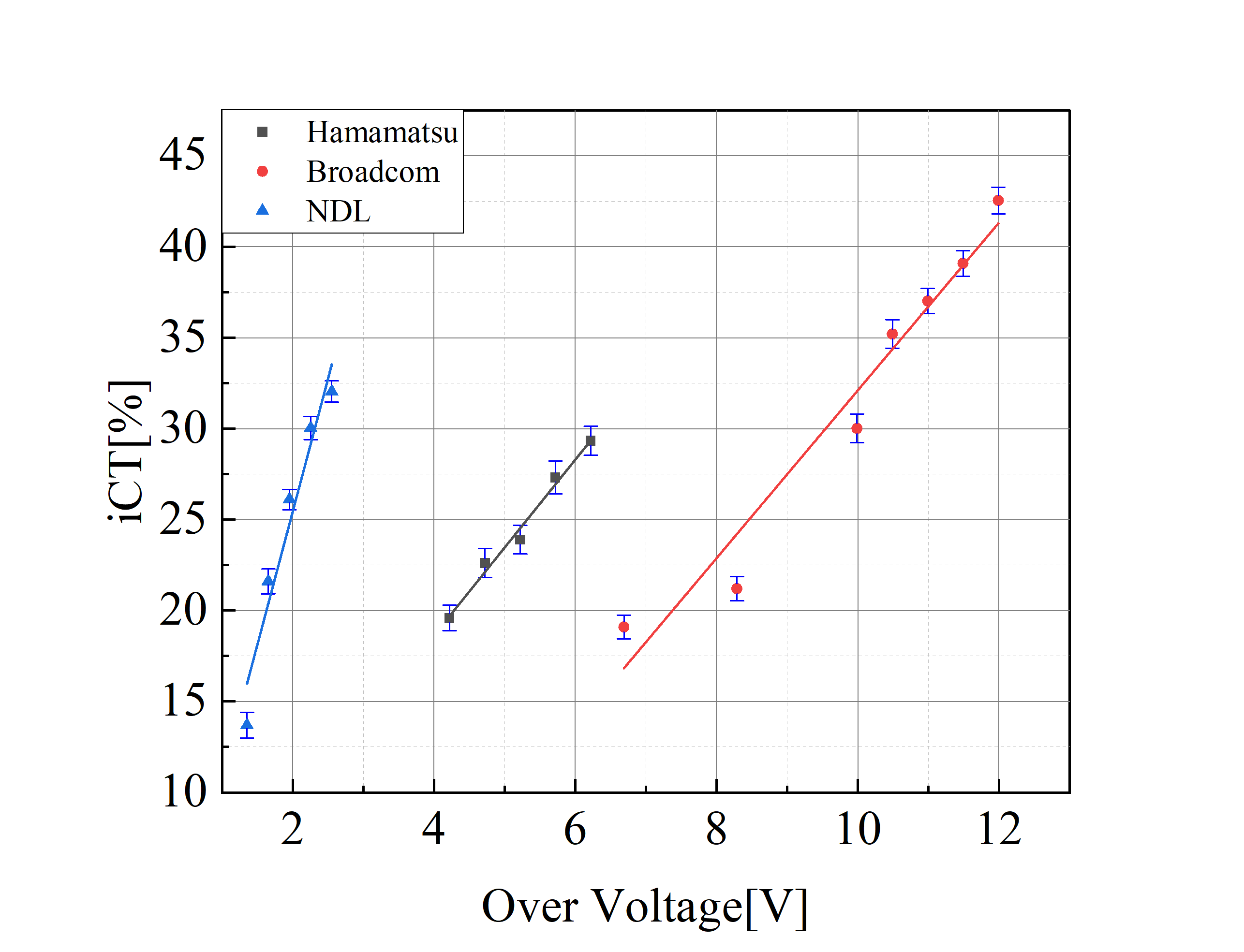}
\caption{The probability of iCT occurring at 40~K for three SiPMs as a function of overvoltage. The error bars on the Y-axis represent statistical errors generated during the calculation of iCT probability.}\label{fig:12}
\end{figure}

The iCT ratio \cite{33} \(P(iCT)\) of a SiPM can be roughly calculated by the following formula:

\begin{equation}
P(iCT)=\frac{N(1.5P.E)}{N(0.5P.E)} 
\end{equation}

\(N(1.5P.E.)\) is the count exceeding the 1.5 P.E. threshold in the dark noise energy spectrum, and \(N(0.5P.E.)\) is the count exceeding the 0.5 P.E. threshold in the dark noise energy spectrum, with both charge and amplitude being less than 2 P.E..

Through probability analysis of iCT for three SiPMs under different overvoltage conditions (fig.~\ref{fig:12}), within a certain bias voltage range, the probability of iCT increases as overvoltage increases.

\subsection{SPE resolution}\label{subsec6}
\begin{figure}[h]
\centering
\includegraphics[width=0.495\textwidth]{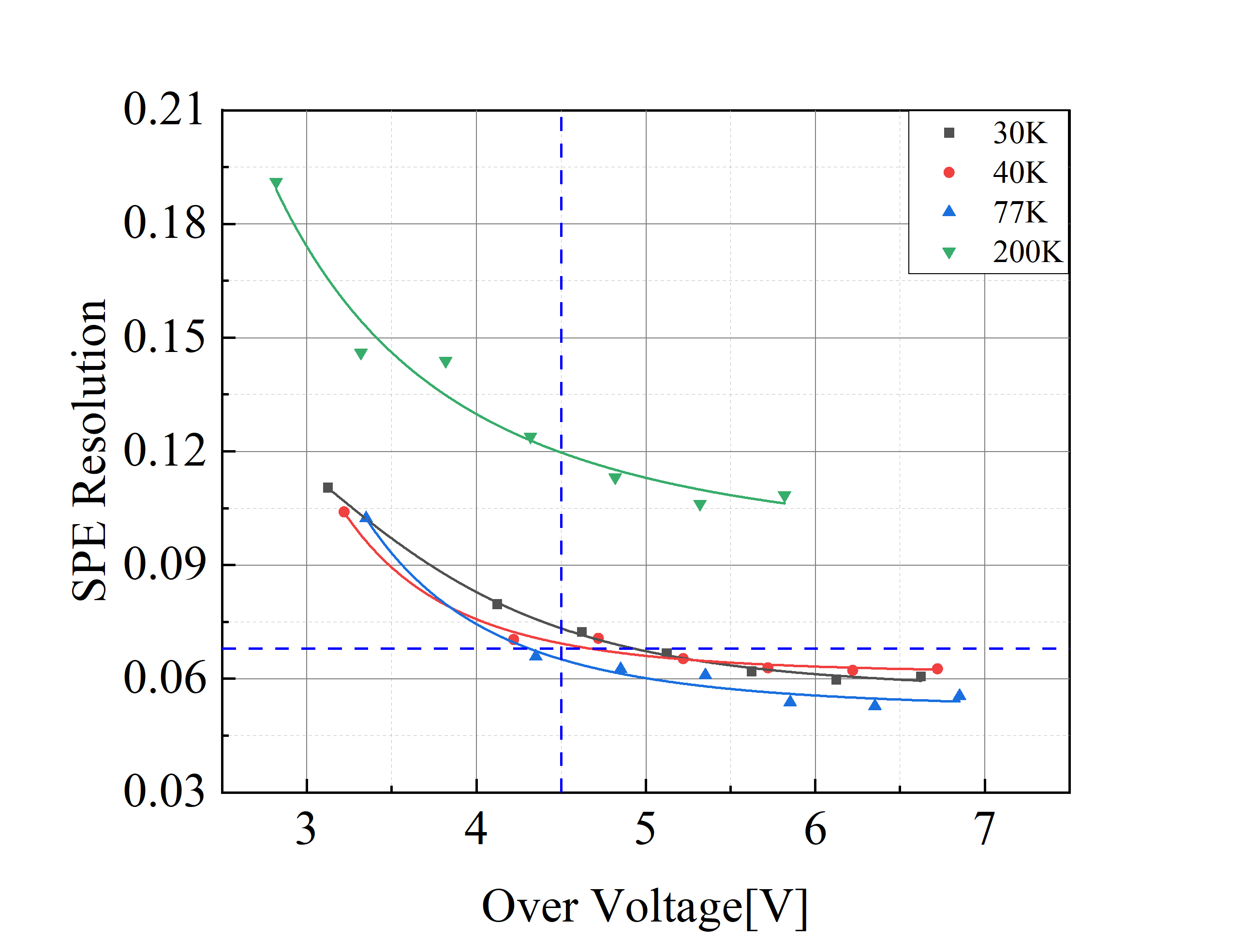}
\includegraphics[width=0.495\textwidth]{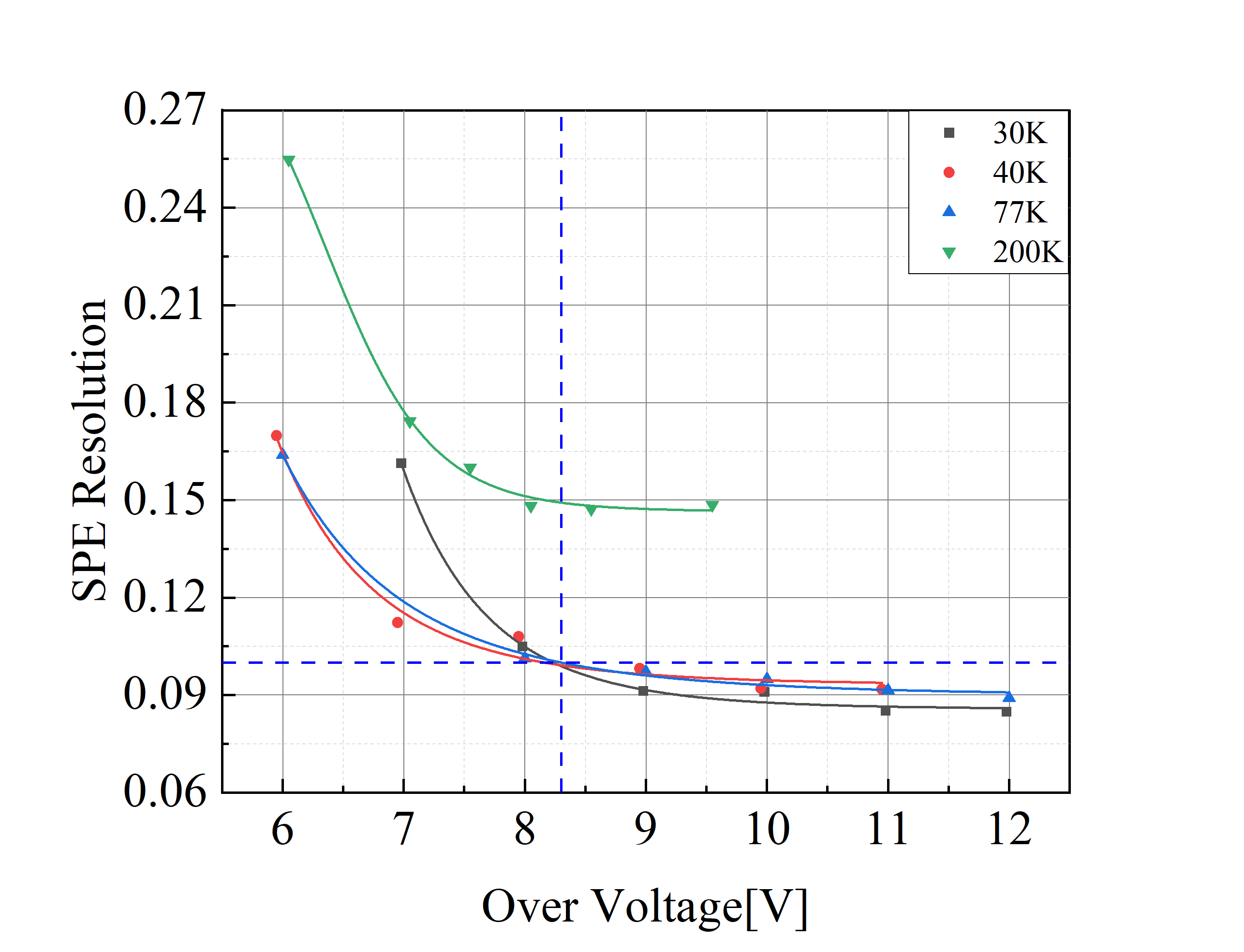}
\includegraphics[width=0.495\textwidth]{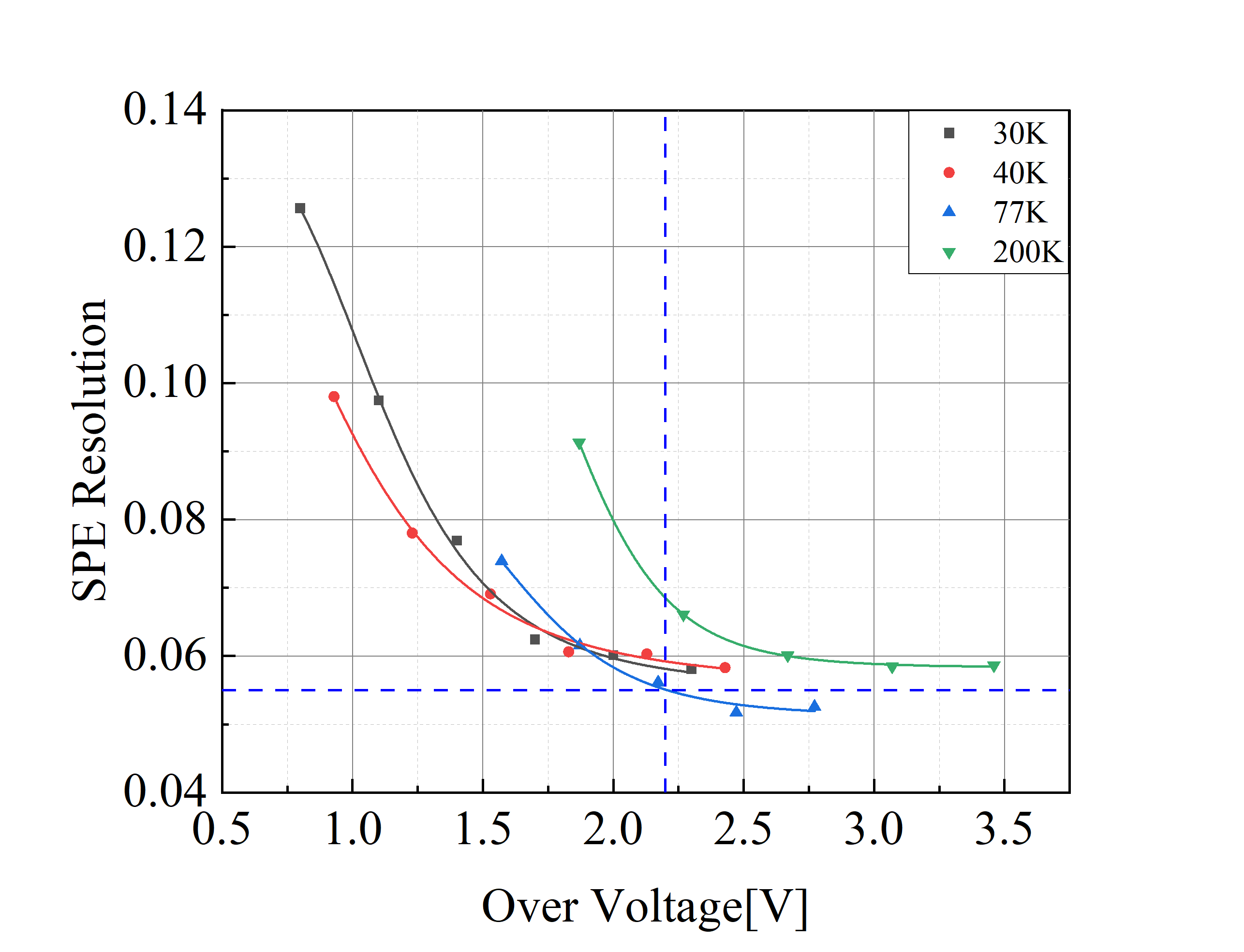}
\includegraphics[width=0.495\textwidth]{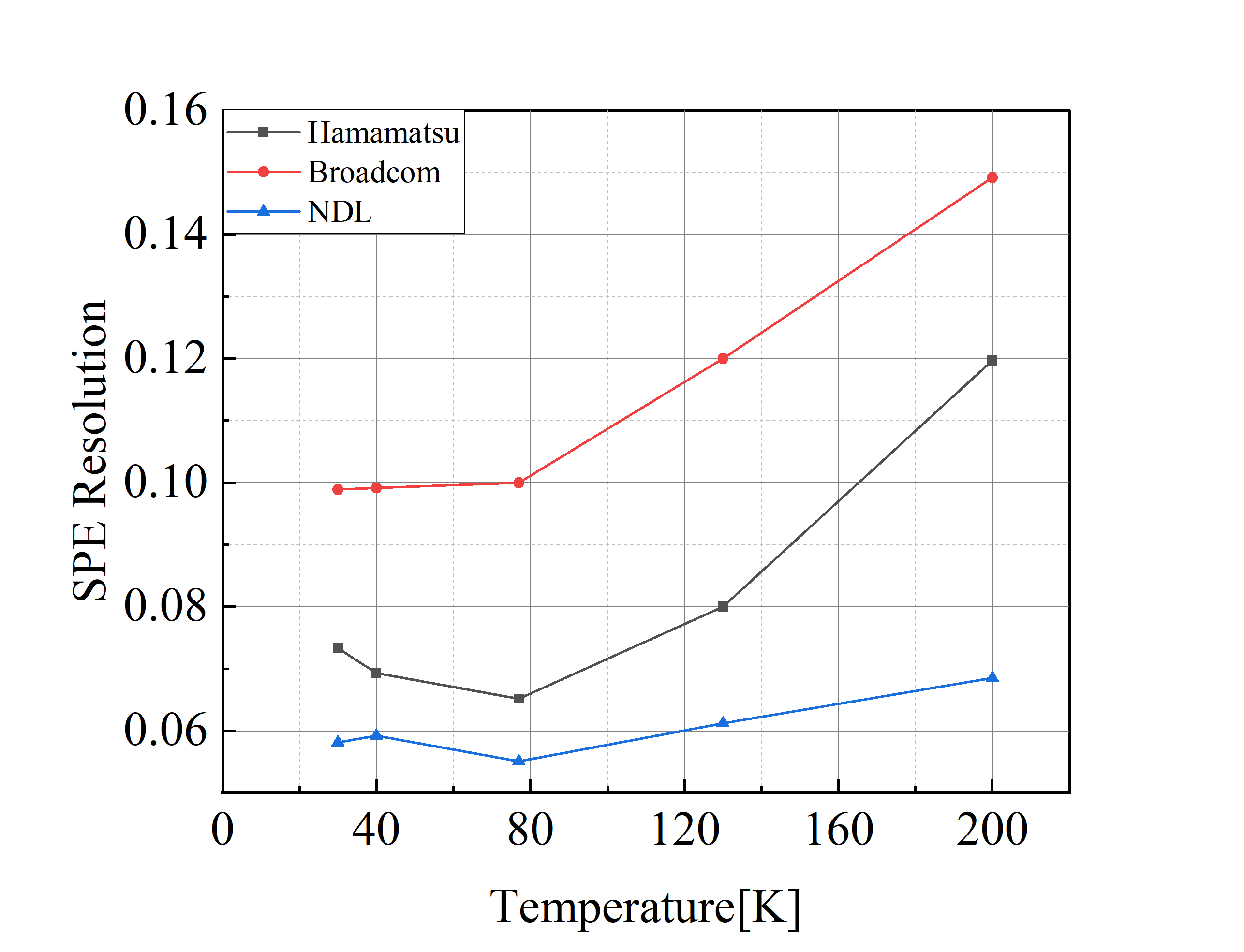}
\caption{The SPE resolution of three SiPMs at different temperatures (30~K, 40~K, 77~K, and 200~K) under varying overvoltage is presented. The upper left figure corresponds to the Hamamatsu SiPM, the upper right figure to the Broadcom SiPM, and the lower left figure to the NDL SiPM. The lower right figure shows the variation of SPE resolution with temperature for the Hamamatsu SiPM, Broadcom SiPM, and NDL SiPM at overvoltage of 4.5~V, 8.3~V, and 2.2~V, respectively. For all three SiPMs, within the low overvoltage range, the SPE resolution improves significantly as the overvoltage increases.}\label{fig:13}
\end{figure}

Against the background of SiPM's dark noise, the $\sigma$/mean ratio \cite{34} of the single-photon energy spectrum can effectively characterize the SPE resolution (fig.~\ref{fig:13}), which intuitively reflects the device's ability to detect single photons. The SPE resolution under different temperature conditions was measured experimentally, and the results showed that temperature changes significantly affect the resolution characteristics. 

Under constant temperature conditions, there is an obvious positive correlation between SPE resolution and overvoltage in the low overvoltage ranges: 3~V to 4.5~V for Hamamatsu SiPM, 6~V to 8.3~V for Broadcom SiPM, and 0.8~V to 2.2~V for NDL SiPM. The small test overvoltage range of NDL SiPM is mainly because, at 40~K, when the overvoltage is slightly too high, the SiPM is prone to secondary breakdown and cannot work normally.

As the bias voltage increases, the combined effects of enhanced avalanche gain and improved signal-to-noise ratio lead to a rapid improvement in SPE resolution. However, when the overvoltage exceeds a critical value, the improvement in resolution tends to saturate and may even slightly deteriorate, constrained by factors such as intensified baseline noise and increased optical crosstalk. This regularity provides important guidance for optimizing the operating parameters of SiPMs. By precisely regulating the operating temperature and optimizing the overvoltage settings, the single-photon resolution of SiPMs can be enhanced.

When the SiPM is combined with a surface-coated TPB (tetraphenyl butadiene) wavelength conversion layer, it shifts the 340 nm \cite{35} photons emitted by the pure-CsI crystal to 420 nm. This wavelength band exactly matches the PDE peak response range of the SiPMs, providing an optimal detection scheme for subsequent light yield experiments.

\section{Conclusion}\label{sec4}
This study systematically characterized the key performance parameters of three SiPMs under different temperature conditions by constructing a low-temperature testing system. The research shows that in a low-temperature operating environment, the gain of SiPMs is improved, the SPE resolution is optimized, and the DCR is reduced by 7 orders of magnitude. All three SiPMs can work normally within the temperature range of 30~K to 293~K. When the Hamamatsu SiPM, Broadcom SiPM, and NDL SiPM operate at 40~K with overvoltages of approximately 4.5~V, 8.3~V, and 2.2~V respectively, the devices exhibit the optimal combination of performance, and the single-photon detection resolution is significantly improved.

Studies have shown that in terms of gain, the Broadcom SiPM exhibits excellent characteristics. The photon signal broadening of the Broadcom SiPM in a low-temperature environment reaches the microsecond level, an order of magnitude higher than that of the other two devices. In terms of SPE resolution, the Hamamatsu SiPM and NDL SiPM perform better. The Broadcom SiPM and NDL SiPM have more advantages in dark noise control, among which the Broadcom SiPM can achieve a DCR of 0.006~\si{\Hz\per\mm^2} under 40~K. The NDL SiPM, due to its small operating voltage range at 40~K, is prone to secondary breakdown, which interferes with the stability of the test. In the normal-temperature operating mode, the Broadcom SiPM shows a higher photon detection efficiency (PDE = 63\% vs. Hamamatsu's 50\%). Based on multi-dimensional performance evaluation, it is recommended to use the Broadcom SiPM to build the array.

The testing method established in this study and the obtained optimal parameter range of SiPM will ensure the reliability and accuracy of the research on the light yield of pure CsI crystals at low temperatures. It is of great significance for promoting the development of low-temperature detector technology.

\section{Acknowledgement}\label{sec5}
This work is supported by the State Key Laboratory of Particle Detection and Electronics (SKLPDE-ZZ-202116), the Natural Science Foundation of Hunan Province (Contract Nos. 2025JJ60063 and 2024JJ2044) and the Research Foundation of Education Bureau of Hunan Province, China (Grant No. 22A0286).

\section{Conflict of interest}\label{sec6}
The authors declare that they have no conflict of interest.

\end{document}